\documentclass[twocolumn,twoside]{IEEEtran}

\usepackage[utf8]{inputenc}

\usepackage{amsmath,amssymb,amsfonts,amsthm,bm}
\usepackage{dsfont}

\usepackage{graphicx}
\usepackage{array}
\usepackage{tabularx}
\usepackage{multirow}
\usepackage{booktabs}
\usepackage{makecell}
\usepackage{float}

\usepackage{xcolor}
\usepackage{url}
\usepackage{cite}

\usepackage{algorithm}
\usepackage{algorithmicx}
\usepackage[noend]{algpseudocode}

\usepackage{acronym}
\usepackage{tablefootnote}
\usepackage{stfloats}
\usepackage{balance}

\usepackage{subfigure}
\usepackage{caption}

\allowdisplaybreaks[4]

\newcommand{\figuresname}{Figs.}
\renewcommand{\tablename}{Table}

\newcommand{\meter}{\,\mathrm{m}}
\newcommand{\mm}{\,\mathrm{mm}}
\newcommand{\cm}{\,\mathrm{cm}}

\newcommand{\dB}{\,\mathrm{dB}}

\newcommand{\THz}{\,\mathrm{THz}}

\acrodef{AWGN}{additive white Gaussian noise}
\acrodef{AdamW}{adaptive moment estimation with weight decay}
\acrodef{B5G}{beyond-5G}
\acrodef{CSA}{central satellite}
\acrodef{CFmMIMO}{cell-free massive \acs{MIMO}}
\acrodef{mMIMO}{massive \acs{MIMO}}
\acrodef{DPD} {disk Poisson Process} 
\acrodef{DRA}{direct radiating antenna}
\acrodef{DC}{direct current}
\acrodef{DoD}{depth of discharge}
\acrodef{eMBB}{enhanced mobile broadband}
\acrodef{E-GPW}{extended gridded population of world database}
\acrodef{FDM}{frequency division multiplexing}
\acrodef{FIR}{finite impulse response}
\acrodef{TFoA}{thinned \acs{FoA}}
\acrodef{FoA}{formation of arrays}
\acrodef{GEO}{geostationary Earth orbit}
\acrodef{GSO}{geosynchronous Earth orbit}
\acrodef{H-RRM}{heuristic radio resource management}
\acrodef{KPI}{key performance indicator}
\acrodef{IF}{intermediate frequency}
\acrodef{LEO}{low Earth orbit}
\acrodef{LoS}{line-of-sight}
\acrodef{MAI}{multiple access interference}
\acrodef{MB}{multi beam}
\acrodef{MEO}{medium Earth orbit}
\acrodef{mMTC}{massive machine-type communications}
\acrodef{eMTC}{enhanced machine-type communications}
\acrodef{PDD}{Poisson disk distribution}
\acrodef{EIRP}{effective isotropic radiated power}
\acrodef{NTN}{non terrestrial network}
\acrodef{URLLC}{ultra-reliable low-latency communications}
\acrodef{ECEF}{Earth-centered Earth-fixed}
\acrodef{GPS}{global positioning system}
\acrodef{HTFS}{high throughput fractionated satellite}
\acrodef{UHF}{ultra-high frequency}
\acrodef{FF}{formation flying}
\acrodef{MIMO}{multiple input multiple output}
\acrodef{SISO}{single input single output}
\acrodef{M-MIMO}{massive multiple input multiple output}
\acrodef{GNC}{guidance navigation and control}
\acrodef{GNSS}{global navigation satellite system}
\acrodef{OFDM}{orthogonal frequency division multiplexing}
\acrodef{PHY}{physical-layer}
\acrodef{SA}{sub-array}
\acrodef{SADM}{solar array drive mechanism}
\acrodef{SG}{solar generator}
\acrodef{SNR}{signal-to-noise ratio}
\acrodef{SIR}{signal-to-interference ratio}
\acrodef{SINR}{signal-to-interference-plus-noise ratio}
\acrodef{SSPA}{solid-state power amplifier}
\acrodef{RF}{radio frequency}
\acrodef{R-GEO}{regional \acs{GEO}}
\acrodef{UT}{user terminal}
\acrodef{UE}{user equipment}
\acrodef{QVD}{quantized virtual distancing}
\acrodef{TDM}{time division multiplexing}
\acrodef{R-GEO}{regional GEO}
\acrodef{RRM}{radio resource management}
\acrodef{SSPA}{solid-state power amplifier}
\acrodef{BLER}{block error rate}
\acrodef{DL}{downlink}
\acrodef{UL}{uplink}
\acrodef{PSD}{power spectral density}
\acrodef{UDSM}{ultra deep sub-micron}
\acrodef{PEPS}{power energy platform simulation}
\acrodef{SE}{spectral efficiency}
\acrodef{wrt}{with respect to}
\acrodef{OBP}{on-board digital processor}
\acrodef{UDSM}{ultra-deep-sub-micron}
\acrodef{3GPP}{third generation partnership project}
\acrodef{AWGN}{additive white Gaussian noise}
\acrodef{B5G}{beyond-5G}
\acrodef{CS}{central satellite}
\acrodef{DRA}{direct radiating antenna}
\acrodef{DC}{direct current}
\acrodef{DoD}{depth of discharge}
\acrodef{eMBB}{enhanced mobile broadband}
\acrodef{FDM}{frequency division multiplexing}
\acrodef{FIR}{finite impulse response}
\acrodef{TFoA}{thinned \acs{FoA}}
\acrodef{FoA}{formation of arrays}
\acrodef{GEO}{geostationary Earth orbit}
\acrodef{GSO}{geosynchronous Earth orbit}
\acrodef{KPI}{key performance indicator}
\acrodef{IF}{intermediate frequency}
\acrodef{LEO}{low Earth orbit}
\acrodef{LoS}{line-of-sight}
\acrodef{MAI}{multiple access interference}
\acrodef{MEO}{medium Earth orbit}
\acrodef{mMTC}{massive machine-type communications}
\acrodef{eMTC}{enhanced machine-type communications}
\acrodef{EIRP}{effective isotropic radiated power}
\acrodef{NTN}{non terrestrial network}
\acrodef{URLLC}{ultra-reliable low-latency communications}
\acrodef{ECEF}{Earth-centered Earth-fixed}
\acrodef{GPS}{global positioning system}
\acrodef{HTFS}{high throughput fractionated satellite}
\acrodef{HTS}{high throughput satellite}
\acrodef{UHF}{ultra-high frequency}
\acrodef{FF}{formation flying}
\acrodef{MIMO}{multiple input multiple output}
\acrodef{GNC}{guidance navigation and control}
\acrodef{NTN}{non-terrestrial network}
\acrodef{GNSS}{global navigation satellite system}
\acrodef{OFDM}{orthogonal frequency division multiplexing}
\acrodef{PHY}{physical-layer}
\acrodef{SA}{satellite array}
\acrodef{SADM}{solar array drive mechanism}
\acrodef{SG}{solar generator}
\acrodef{SNR}{signal-to-noise ratio}
\acrodef{SIR}{signal-to-interference ratio}
\acrodef{SINR}{signal-to-interference-plus-noise ratio}
\acrodef{SSPA}{solid-state power amplifier}
\acrodef{RF}{radio frequency}
\acrodef{R-GEO}{regional \acs{GEO}}
\acrodef{UT}{user terminal}
\acrodef{UE}{user equipment}
\acrodef{TDM}{time division multiplexing}
\acrodef{RRM}{radio resource management}
\acrodef{SSPA}{solid-state power amplifier}
\acrodef{BLER}{block error rate}
\acrodef{DL}{down-link}
\acrodef{UL}{up-link}
\acrodef{RV}{random variable}
\acrodef{PSD}{power spectral density}
\acrodef{UDSM}{ultra deep sub-micron}
\acrodef{PEPS}{power energy platform simulation}
\acrodef{SE}{spectral efficiency}
\acrodef{wrt}{with respect to}
\acrodef{OBP}{on-board digital processor}
\acrodef{UDSM}{ultra-deep-sub-micron}
\acrodef{UC}{user-centric}
\acrodef{CSI}{channel state information}
\acrodef{PAC}{per-antenna constraint}
\acrodef{FPAC}{fair \acs{PAC}}
\acrodef{MPC}{maximum power constraint}
\acrodef{AWGN}{additive white Gaussian noise}
\acrodef{TX}{transmit}
\acrodef{MMSE}{minimum mean-square error}
\acrodef{MF}{matched filter}
\acrodef{ZF}{zero forcing}
\acrodef{ELSA}{enhanced logarithmic spiral array}
\acrodef{UPA}{uniform planar array}
\acrodef{NB}{narrowband}
\acrodef{WB}{wideband}
\acrodef{BFN}{beamforming network}
\acrodef{MD-MIQP}{minimum distance mixed integer quadratic problem}
\acrodef{PDF}{probability density function}
\acrodef{NPR}{noise-to-power ratio}
\acrodef{HPA}{high power amplifier}
\acrodef{CF}{cell-free}
\acrodef{3GPP}{third generation partnership project}
\acrodef{UC-MIMO}{user centric MIMO}
\acrodef{VHTS}{very high throughput satellites}
\acrodef{PM-MIMO}{pragmatic M-MIMO}
\acrodef{NR}{new radio}
\acrodef{LMS}{land mobile satellite}
\acrodef{IM}{intermodulation}
\acrodef{OBO}{output back-off}
\acrodef{FDD}{frequency division duplexing}
\acrodef{TDD}{time division duplexing}
\acrodef{BLER}{block error rate}
\acrodef{rv}{random variable}
\acrodef{COB}{center of beam}
\acrodef{ISL}{inter-satellite link}
\acrodef{AP}{access point}
\acrodef{SVD}{single value decomposition}
\acrodef{CW}{continuous wave}
\acrodef{AOCS}{attitude and orbit control system}
\acrodef{LSTM}{long short-term memory}
\acrodef{CPU}{central processing unit}
\acrodef{GPU}{graphics processing unit}
\acrodef{TPU}{tensor processing unit}
\acrodef{MMF}{max-min fairness}
\acrodef{MHA}{multi-head attention}
\acrodef{FFN}{feed-forward network}
\acrodef{MSE}{mean-square error}
\acrodef{CDF}{cumulative distribution function}
\acrodef{EPA}{equal power allocation}
\acrodef{FPA}{fractional power allocation}
\acrodef{TNN}{transformer neural network}
\acrodef{THz}{teraHertz}
\acrodef{RIS}{reconfigurable intelligent surface}
\acrodef{NF}{near-field}
\acrodef{FF}{far-field}
\acrodef{LMMSE}{linear minimum mean-square error}
\acrodef{EMI}{electromagnetic interference}
\acrodef{BS}{base station}
\acrodef{NMSE}{normalized mean-square error}
\acrodef{SNR}{signal-to-noise ratio}
\acrodef{SIR}{signal-to-interference ratio}
\acrodef{AO}{alternating optimization}
\acrodef{DoA}{direction of arrival}
\acrodef{DoAs}{directions of arrival}
\acrodef{PG}{projected gradient}
\acrodef{RS-LS}{reduced-subspace least squares}
\acrodef{MISO}{multiple-input single-output}

\begin{document}

\title{\LARGE{Near-Field \acs{MMSE} Channel Estimation for \acs{THz} \acs{RIS}-aided Communications with Electromagnetic Interference}}

\author{Wen-Xuan Long, \emph{Member, IEEE}, Marco Moretti, \emph{Member, IEEE},\\ Giacomo Bacci, \emph{Senior Member, IEEE} and Luca Sanguinetti, \emph{Fellow, IEEE}

\thanks{W.-X. Long, M. Moretti, G. Bacci and L. Sanguinetti are with the Dipartimento di Ingegneria dell’Informazione, University of Pisa, 56126 Pisa, Italy (e-mail: wenxuan.long@ing.unipi.it, marco.moretti@unipi.it, giacomo.bacci@unipi.it, luca.sanguinetti@unipi.it).}
\thanks{M. Moretti, G. Bacci and L. Sanguinetti are also with National Inter-University Consortium for Telecommunications (CNIT), 43124 Parma, Italy.}
\thanks{This work has been performed in the framework of the HORIZON-JU-SNS-2022 EU project TIMES under grant no. 101096307, by the European Union under the Italian National Recovery and Resilience Plan (NRRP) of NextGenerationEU, partnership on “Telecommunications of the Future” (PE00000001 – Program ``RESTART'', Cascade Call SMART), and also supported in part by Italian Ministry of Education and Research (MUR) in the Framework of the FoReLab Project (Departments of Excellence) and the Project GARDEN funded by EU in NextGenerationEU Plan through Italian ``Bando Prin 2022-D.D.1409 del 14-09-2022''.}
}
\maketitle

\begin{abstract}
This letter investigates the channel estimation problem in \ac{THz} wireless communications where a \ac{RIS} is employed to assist wireless transmission between different devices. Unlike existing studies, we consider a novel scenario where specific devices are all located in the radiative \ac{NF} region of the \ac{RIS}. Meanwhile, we also account for the impact on channel estimation of the random electromagnetic interference occurring at the \ac{RIS} location. A linear minimum mean-square error estimator is employed, where the estimation error is fully determined by the \ac{RIS} configuration. Optimizing the \ac{RIS} involves solving a non-convex problem, which is addressed using an alternating
optimization approach based on the diagonally scaled gradient descent algorithm. Numerical results in the \ac{THz} band highlight the importance of leveraging NF channel statistics over far-field approximations and demonstrate that the proposed estimator achieves substantial improvements in normalized mean-square error compared to existing methods.
\end{abstract}

\begin{IEEEkeywords}
TeraHertz wireless communications, reconfigurable intelligent surfaces, near-field channel estimation, electromagnetic interference, spatially correlated channels.
\end{IEEEkeywords}

\acresetall

\vspace{-0.2cm}
\section{Introduction}

\ac{THz} wireless communications ($0.1-10\THz$) are gaining attention for applications such as space communications, biomedical sensing, and industrial automation, due to their vast spectral resources and ultra-low latency potential \cite{Chen2021Terahertz}. However, their short wavelengths lead to high directivity \cite{Abbasi2020Channel} and severe propagation losses \cite{bodet2024sub,SELIMIS2024102453}, making \ac{LoS} links essential in most scenarios. As a solution, \acp{RIS} have emerged as a promising technology \cite{Long2021A}. By tuning the impedance of passive elements, an \ac{RIS} can redirect signals to form virtual \ac{LoS} links, helping to overcome blockages. Since \acp{RIS} operate passively without amplification, achieving sufficient \ac{SNR} typically requires a large reflective surface.


As the \ac{RIS} size grows and the wavelength shortens, the Fraunhofer distance \cite{Liu2023Near}, which separates the radiative \ac{NF} and \ac{FF} regions, significantly increases in \ac{RIS}-aided \ac{THz} systems. Consequently, communication devices are likely to lie within the radiative \ac{NF} region, where spherical wavefronts induce distance- and angle-dependent phase variations across \ac{RIS}. These characteristics fundamentally alter the channel model, necessitating a reassessment of conventional \ac{FF} channel estimation methods for \ac{NF} scenarios.

Furthermore, due to the high path loss and noise sensitivity of \ac{RIS}-aided \ac{THz} communication systems, they are particularly vulnerable to interference \cite{guo2023robust,Middleton1977Statistical}, including both intentional \cite{Guerboukha2021Jamming,Shrestha2022Jamming} and unintentional \cite{Petrov2017Interference} sources. Existing studies on \ac{EMI}-robust cascaded channel estimation are mainly limited to \ac{RIS}-aided \ac{FF} systems in low-frequency bands, where \ac{RS-LS} \cite{Long2023Channel} and \ac{LMMSE} \cite{Long2024MMSE} estimators are used to handle constant and random interference, respectively. In contrast, \ac{EMI} is largely ignored in recent works on \ac{RIS}-aided \ac{NF} \ac{THz} systems (e.g., \cite{Pan2023IRS,Wu2023Parametric}), and to the best of our knowledge, no benchmark scheme explicitly addresses \ac{EMI}-aware channel estimation in \ac{RIS}-aided \ac{THz} \ac{NF} systems.


To address the above challenges, this paper extends the \ac{LMMSE} estimator and \ac{AO} strategy from \cite{Long2024MMSE} to a novel \ac{THz} communication scenario in which specific communication devices and interference sources are all located within the \ac{NF} region of the \ac{RIS}. In this setup, the signal transmitted by a specific device, as well as the random interference, both reach the \ac{RIS} via \ac{NF} \ac{LoS} paths with a small solid angle and are then reflected toward another specific device. The resulting channels and \ac{EMI} are jointly characterized by the distances and \acp{DoA} from the node to the \ac{RIS}, unlike far-field channels determined solely by \ac{DoA} \cite{Long2024MMSE}. Leveraging these \ac{NF} statistical structures, we estimate the uplink cascaded channel using the \ac{LMMSE} estimator from \cite{Long2024MMSE}, and optimize the \ac{RIS} phase-shifts via the corresponding two-step \ac{AO} algorithm. Simulations in the \ac{THz} band confirm the necessity of employing \ac{NF} statistical knowledge in this setting and show that the proposed \ac{NF}-aware \ac{LMMSE} approach significantly outperforms existing benchmarks in terms of \ac{NMSE}.

\vspace{-0.2cm}
\section{System Model}

We consider a \ac{RIS}-aided system operating in the \ac{THz} band, where two specific single-antenna devices communicate with each other via a \ac{RIS}, in a scenario similar to the one depicted in \cite[\figurename~1]{Long2024MMSE} and not reported here for the sake of brevity. The \ac{RIS} comprises $N = N_H \times N_V$ passive elements arranged as a \ac{UPA} configuration, with inter-element spacing $d_N$ in both the horizontal and vertical directions. The \ac{RIS} is positioned on the $yoz$-plane in a three-dimensional coordinate system, as illustrated in \cite[\figurename~1]{Emil2021Rayleigh}.

The Fraunhofer distance of the \ac{RIS}, given by $R = 2D^2/\lambda$ \cite{Liu2023Near}, defines the boundary between the \ac{FF} and \ac{NF} regions, where $D = \sqrt{N_H^2 + N_V^2}d_N$ is the aperture size and $\lambda$ is the wavelength. As the aperture grows, often spanning tens or hundreds of wavelengths, this distance increases significantly. For instance, a $0.2 \times 0.1\meter^2$ \ac{RIS} with two-wavelength inter-element spacing at $0.3\THz$ ($\lambda = 1\mm$) has $N = 5000$ elements and a Fraunhofer distance of $R = 100\meter$, covering much of a typical cell. Thus, communication devices likely lie within the \ac{NF} region, invalidating the \ac{FF} assumption and calling for a more accurate \ac{NF} model.


\vspace{-0.2cm}
\subsection{NF channel model}

Following \cite{Abbasi2020Channel}, \ac{THz} \ac{NF} channels assume the signal sent by the specific device arrives at \ac{RIS} within a narrow solid angular spread centered around the \ac{LoS} direction. For a specific device at distance $r_h$, with DoA $(\varphi_h, \theta_h)$, the channel vector $\mathbf{h} \in \mathbb{C}^N$ follows a correlated Rayleigh fading model $\mathbf{h} \sim \mathcal{N}_{\mathbb{C}}(\mathbf{0}_N, \mathbf{R}_h)$, defined by the spatial correlation matrix \cite{LONG2025}:
\vspace{-0.1cm}
\begin{align}\label{RhNF}
\mathbf{R}_h=& \beta_h
\int_{ r_h-\Delta^{(r)}_h}^{ r_h+\Delta^{(r)}_h}\!\!
\int_{\varphi_h -\Delta^{(\varphi)}_h}^{\varphi_h +\Delta^{(\varphi)}_h}\!\!
\int_{ \theta_h -\Delta^{(\theta)}_h}^{ \theta_h +\Delta^{(\theta)}_h}\!\!
f_{h}( r, \varphi , \theta )\cdot \nonumber\\&
\qquad\mathbf{a}( r, \varphi , \theta )\mathbf{a}^{\mathrm{H}}( r, \varphi
, \theta )d \theta d\varphi d r,
\end{align}
\vspace{-0.1cm}
where $\beta_h = \frac{1}{N} \mathrm{tr}\{\mathbf{R}_h\}$ is the average channel power, and $(\Delta^{(r)}_h, \Delta^{(\varphi)}_h, \Delta^{(\theta)}_h)$ denote the distance and angular spreads. The function $f_h(\cdot)$ is the normalized spatial scattering function \cite{Demir2024Spatial}, and $\mathbf{a}({r}, {\varphi}, {\theta})$ is the array response vector, defined as
\begin{align}
\mathbf{a}( r, \varphi , \theta )=\left[1,\ldots,e^{j\frac{2\pi }{\lambda
}( r_{n}- r)},\ldots ,e^{j\frac{2\pi }{\lambda }( r_{N}- r)}\right]^{
\text{T}},
\label{array_response}
\end{align}%
where $r_{n}$ and $ r$ denote the distances from a point within the spatial spread region to the $n$-th \ac{RIS} element and to the reference element, and $(\varphi,\theta)$ are the \ac{DoA} of the point in the spatial spread region relative to the RIS. The distance $r_{n}$ is computed as \cite{LONG2025}
\vspace{-0.1cm}
\begin{equation}
\label{rmU}
 r_{n}= r\sqrt{1-\frac{2\mathbf{k}^{\text{T}}( \varphi ,\theta )\mathbf{u}_{n}%
}{ r}+\frac{\Vert \mathbf{u}_{n}\Vert ^{2}}{ r^{2}}},
\end{equation}
where $\mathbf{k}( \varphi , \theta ) = [\cos {\theta }\cos {\varphi },\cos {%
\theta }\sin {\varphi },\sin {\theta }]^{\text{T}}$ is the wave vector from the point in the spatial spread region to the \ac{RIS}, and $\mathbf{u}_{n}$ denotes the coordinate of the $n$-th \ac{RIS} element.


We then define the vector $\mathbf{g}\in \mathbb{C}^N$ as the channel between the \ac{RIS} and the other specific receiving device. It also follows a correlated Rayleigh fading model, i.e. $\mathbf{g} \sim \mathcal{N}_{\mathbb{C}}(\mathbf{0}_N, \mathbf{R}_g)$, where $\mathbf{R}_g$ denotes the corresponding spatial correlation matrix.
The channels $\mathbf{g}$ and $\mathbf{h}$ are statistically independent. Hence, the cascaded channel between the two specified devices via the \ac{RIS} is given by
\vspace{-0.1cm}
\begin{align}
\mathbf{x} = \mathbf{g} \odot \mathbf{h},
\end{align}
\vspace{-0.1cm}\noindent
where $\odot$ represents the Hadamard product.

\vspace{-0.2cm}
\subsection{NF EMI model}

Given the high path loss inherent to \ac{THz} channels, the \ac{EMI}, whether resulting from unintended signals emitted by other devices \cite{Petrov2017Interference} or from maliciously injected high-power electromagnetic signals \cite{Guerboukha2021Jamming, Shrestha2022Jamming}, exerts a non-negligible impact on \ac{THz} wireless communications. Thus, in this paper, we additionally consider $\mathbf{e}(i) \in \mathbb{C}^N$, which is the \ac{EMI} incident on the \ac{RIS} at the $i$-th channel use during the training phase. It is modeled as $\mathbf{e}(i) \sim \mathcal{N}_{\mathbb{C}}(\mathbf{0}_N, \sigma_e^2 \mathbf{R}_e)$ with independent realizations across uses, i.e., $\mathbb{E}\{\mathbf{e}(i)\mathbf{e}(i')^{\text{H}}\} = 0$ for $i \neq i'$. The normalized spatial correlation matrix $\mathbf{R}_e$ shares the structure in \eqref{RhNF}, but with a different spatial scattering function $f_e({r}, {\varphi}, {\theta})$.

\section{Channel Estimation and \acs{RIS} Optimization}

\subsection{Pilot Transmission and LMMSE Estimation}

We consider a system where $\tau$ channel uses are allocated for uplink channel estimation before data transmission. Since the \ac{RIS} typically serves to assist communication between two devices that do not have a direct link, the channel between the two specific devices is neglected. An all-one training sequence of length $\tau$ is used. Let $\bm{\phi}(i) \in \mathbb{C}^{N}$ collect the adjustable \ac{RIS} phase-shifts $\{\phi_n(i) \in [0, 2\pi); n = 1, \ldots, N\}$ at channel use $i$. The received training signal at the $i$-th channel use is
\vspace{-0.1cm}
\begin{equation}\label{ym2}
y_{\mathrm{tr}}(i)=\sqrt{\rho^{\mathrm{tr}}}\bm{\phi}(i)^{\mathrm{T}}\mathbf{x}+w(i)+z(i),
\end{equation}
where
\begin{equation}
w(i) = \bm{\phi}(i)^{\mathrm{T}}\left(\mathbf{g} \odot\mathbf{e}(i)\right)
\end{equation}
is the \ac{EMI} reflected from the \ac{RIS} to the specific receiving device, $\rho^{\mathrm{tr}}$ is the power of the training signal, and $z(i)\sim \mathcal{N}_{\mathbb{C}}(0,\sigma_z^2)$ is thermal noise. Collecting the received signals over all $\tau$ channel uses yields the vector \cite{Long2024MMSE}
\begin{equation}
\label{y2.2}
\mathbf{y}_{\mathrm{tr}}=\sqrt{\rho^{\mathrm{tr}}}\mathbf{\Phi}_{\tau}\mathbf{x}+\mathbf{w}_{\mathrm{tr}}+\mathbf{z}_{\mathrm{tr}}
\end{equation}
where $\mathbf{w}_{\mathrm{tr}} = {[w(1), \cdots, w(\tau)]}^{\mathrm{T}}$, $\mathbf{z}_{\mathrm{tr}} = {[z(1), \cdots, z(\tau)]}^{\mathrm{T}}$ and
\begin{equation}
 \mathbf{\Phi}_{\tau} = {[\bm{\phi}(1),\bm{\phi}(2),\cdots,\bm{\phi}(\tau)]}^{\mathrm{T}}\in \mathbb{C}^{\tau \times N}.
\end{equation}
Since the \ac{RIS} is capable of acquiring the statistical properties of the channel and \ac{EMI} through its sensing mode, the second-order statistics of the cascaded channel and \ac{EMI} are assumed to be known \cite{Long2024MMSE}, and are given by
\begin{equation}\label{Rx_prime}
\mathbf{R}_{x}=\mathbf{R}_{g}\odot\mathbf{R}_{h}
\end{equation}
and
\begin{equation}\label{Rw_prime}
\mathbf{R}_{w}^{\mathrm{tr}} =  \left(\mathbf{\Phi}_{{\tau}}\mathbf{R}_{q}\mathbf{\Phi}_{{\tau}}^{\mathrm{H}}\right)\odot\mathbf{I}_{\tau},
\end{equation}
where $\mathbf{R}_{q} = \mathbf{R}_{g}\odot\mathbf{R}_{e}$. The \ac{LMMSE} estimate of $\mathbf{x}$ based on the observation of $\mathbf{y}_{\mathrm{tr}}$ is given by \cite{Long2024MMSE}
\begin{equation}\label{LMMSE_estimator}
\hat{\mathbf{{x}}} = \frac{1}{\sqrt{\rho^{\mathrm{tr}}}}\mathbf{\Lambda}(\mathbf{\Phi}_{{\tau}})\mathbf{y}_{\mathrm{tr}},
\end{equation}
\noindent where
\begin{equation}\label{Lambda_Phi}
\mathbf{\Lambda}(\mathbf{\Phi}_{\tau}) =\mathbf{R}_{x}\mathbf{\Phi}_{\tau}^{\mathrm{H}}\left({{\bf R}_y^{\mathrm{tr}}}\right)^{-1},
\end{equation}
\noindent and
\begin{equation}\label{R_y_tr}
{{\bf R}_y^{\mathrm{tr}}} =\mathbf{\Phi}_{\tau}\mathbf{R}_{x}\mathbf{\Phi}_{\tau}^{\mathrm{H}}+\frac{\sigma^{2}_{e}}{\rho^{\mathrm{tr}}}\mathbf{R}_{w}^{\mathrm{tr}}\!+\!\frac{\sigma_z^2}{\rho^{\mathrm{tr}}}\mathbf{I}_{\tau}
\end{equation}
denotes the covariance matrix of $\mathbf{y}_{\mathrm{tr}}$. The \ac{MSE} of $\hat{\mathbf{{x}}}$ takes the form
\begin{equation}\label{MSE_MMSE}
\mathcal{E}_x(\mathbf{\Phi}_{{\tau}}) =\text{tr}\left\{\mathbf{R}_{x}-\mathbf{\Lambda}(\mathbf{\Phi}_{{\tau}})\mathbf{\Phi}_{\tau}\mathbf{R}_x\right\},
\end{equation}
which depends solely on the \ac{RIS} configuration $\mathbf{\Phi}_{{\tau}}$.

\vspace{-0.2cm}
\subsection{\ac{AO}-based \ac{RIS} Optimization}

The minimization of \ac{MSE} \eqref{MSE_MMSE} can be formulated as the following non-convex optimization problem:
\begin{align}
\begin{aligned} \label{opt_1}
&\!\min_{\mathbf{\Phi}_{\tau}\in\mathcal{F}}\ \mathcal{E}_x(\mathbf{\Phi}_{{\tau}})
\end{aligned}
\end{align}
where the feasible set is given by
\begin{align}
\mathcal{F}=\{\mathbf{\Phi}_{\tau} \in \mathbb{C}^{\tau\times N}\vert \ |[\mathbf{\Phi}_{\tau}]_{i,n}|=1;\forall i,n\}
\end{align}
enforcing the unit-modulus constraint on each element of $\mathbf{\Phi}_{\tau}$ due to the passive nature of the \ac{RIS}. Directly solving \eqref{opt_1} is challenging due to its non-convex nature. To address this, \cite{Long2024MMSE} adopts a simplifying assumption by neglecting the dependence of $\mathbf{\Lambda}(\mathbf{\Phi}_{\tau})$ on $\mathbf{\Phi}_{\tau}$. Under this assumption, a two-step \ac{AO} algorithm is employed to iteratively minimize $\mathcal{E}_x$ with respect to $\mathbf{\Lambda}$ and $\mathbf{\Phi}_{\tau}$. In each iteration, the phase-shifts $\mathbf{\Phi}_{\tau}$ from the previous iteration are substituted into \eqref{Lambda_Phi} to update the \ac{LMMSE} estimator $\mathbf{\Lambda}$, which is a direct application of minimizing $\mathcal{E}_x$; subsequently, fixing $\mathbf{\Lambda}$, the phase-shifts $\mathbf{\Phi}_{\tau}$ are updated via a diagonally scaled steepest descent method, as proposed in \cite{Long2024MMSE}. These two steps are repeated alternately until convergence. Due to space limitations, detailed derivations of the \ac{AO} algorithm are omitted. Notably, this algorithm designed for the training phase operates solely based on the known \ac{NF} correlation matrices $\mathbf{R}_x$ and $\mathbf{R}_{w}^{\mathrm{tr}}$, without requiring any online information. Consequently, the optimal \ac{RIS} phase-shifts can be precomputed offline, incurring no additional pilot overhead.


\begin{table}[t!]
\renewcommand{\arraystretch}{1.3}
\setlength{\tabcolsep}{3pt}
\centering
\caption{System parameters.}
\begin{footnotesize}
\begin{tabular}{c|c}
\hline
{ \bf Parameter} & {\bf Value} \\
\hline
Carrier frequency & $f_0 = 0.3\THz$ \\
\hline
Wavelength  & $\lambda = 1\mm$  \\
\hline
Number of \acs{RIS} element & $N_H \times N_V = 12 \times 2$ \\
\hline
Inter-element spacing & $d_N = 10 \lambda=1\cm$ \\
\hline
\acs{SNR} & $0\dB$ \\
\hline
\acs{SIR} & $5\dB$ \\
\hline
Pilot length & $\tau = \text{rank}\{\mathbf{R}_{g_m}\odot\mathbf{R}_{h}\} = 15$ \\
\hline
Transmitter position & $(r_h,\!\varphi_h,\!\theta_h) = (15\meter,\!70^{\circ},\!-20^{\circ})$ \\
\hline
Receiver position & $(r_h,\!\varphi_h,\!\theta_h) = (20\meter,\!-60^{\circ},\!-30^{\circ})$ \\
\hline
\acs{EMI} position & $(r_h,\!\varphi_h,\!\theta_h) = (25\meter,\!-10^{\circ},\!20^{\circ})$ \\
\hline
Angular spread\tablefootnote{Distance and azimuth spreads of entity $x$ are computed using $\Delta^{(r)}_x= r_x/2[\cos(\theta_x+\Delta^{(\theta)}_x)-\cos(\theta_x-\Delta^{(\theta)}_x)]$ and
$\Delta^{(\varphi)}_x=\tan^{-1}\{\Delta^{(r)}_x/(r_x\cos\theta_x)\}$.} of $\mathbf{h}$ & $\Delta^{(\theta)}_h = 1^{\circ}$ \\
\hline
Angular spread of $\mathbf{g}_m$& $\Delta^{(\theta)}_g = 1^{\circ}$ \\
\hline
Angular spread of $\mathbf{e}(i)$& $\Delta^{(\theta)}_e = 3^{\circ}$ \\
\hline
\end{tabular}
\end{footnotesize}
\label{TableI}
\end{table}

\section{Numerical Simulations}

We now evaluate the effectiveness of the \ac{LMMSE} estimator with \ac{AO} in \ac{NF} cascaded channel estimation at \ac{THz} bands, and validate the necessity of incorporating \ac{NF} statistical characteristics in this scenario. Unless stated otherwise, the proposed system operates at $f_0=0.3\THz$, corresponding to $\lambda=1\mm$. The \ac{RIS} is equipped with a $12\times 2$ \ac{UPA} with $d_N =10\lambda$. This configuration results in a Fraunhofer distance $R=29.6\meter$. The specific transmitting and receiving devices are located at $(15\meter,\!70^{\circ},\!-20^{\circ})$ and $(20\meter,\!-60^{\circ},\!-30^{\circ})$ relative to the \ac{RIS}, respectively, and both are located within the \ac{NF} region of the \ac{RIS}. The random \ac{EMI} is generated at $(25\meter,\!-10^{\circ},\!20^{\circ})$ relative to the \ac{RIS}. Other simulation parameters are provided in \tablename~\ref{TableI}. Three different channel estimators are compared:
\begin{enumerate}
    \item The first method, labeled as `RS-LS', is the \ac{RS-LS} estimator for \ac{NF} channel estimation, as defined in \cite[Eq.~(13)]{Demir2024Spatial}. The optimal \ac{RIS} phase-shift for this estimator can be designed according to \cite[Eq.~(26)]{Long2023Channel}. This phase-shift is also used as the initial \ac{RIS} configuration $\mathbf{\Phi}^{(0)}_{\tau}$ in the \ac{AO} strategy;
    \item  The second method is the \ac{LMMSE} estimator with the \ac{AO} strategy employed in this work, labeled as `LMMSE';
     \item To evaluate the gain introduced by the \ac{AO} strategy, we also include a baseline \ac{LMMSE} estimator that directly uses $\mathbf{\Phi}^{(0)}_{\tau}$ without executing the \ac{AO} method. This estimator is labeled as `$\text{LMMSE}$-${\mathbf{\Phi}_0}$'.
\end{enumerate}

\begin{figure}[t!]
\setlength{\abovecaptionskip}{-0.2cm}
\setlength{\belowcaptionskip}{-0.4cm}
\begin{center}
\includegraphics[width = 7.9 cm]{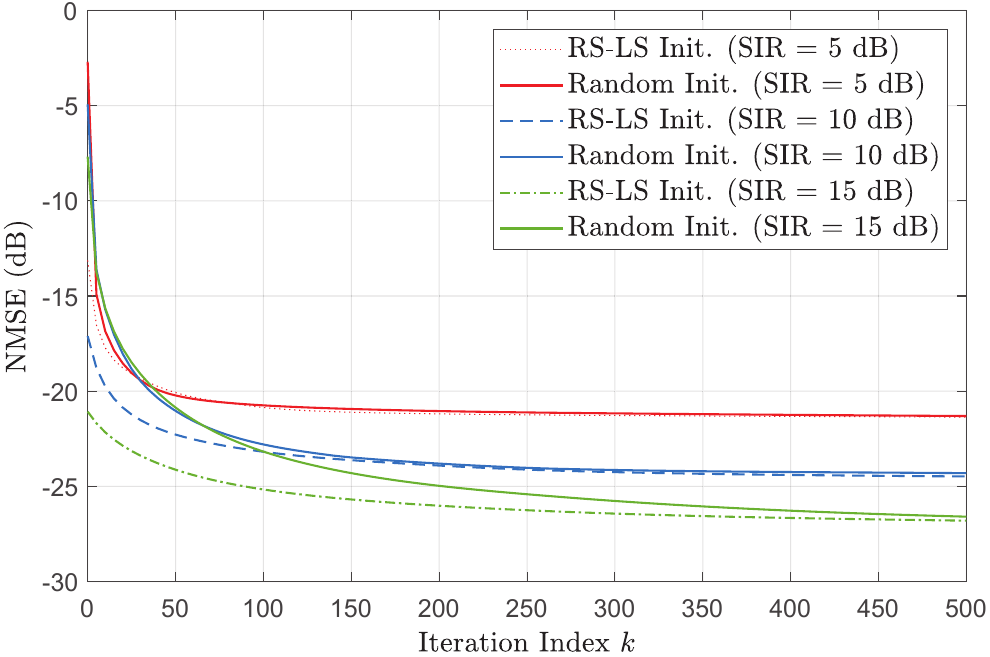}
\end{center}
\caption{\acs{NMSE} vs. iteration index $k$ for different initializations.}
\label{Fig7}
\end{figure}

\begin{figure}[t!]
\setlength{\abovecaptionskip}{-0.3cm}
\setlength{\belowcaptionskip}{-0.4cm}
\begin{center}
\subfigure[Carrier frequency $f_0=0.3\THz$.]{\captionsetup{skip=-0.8cm}\includegraphics[width=7.9 cm]{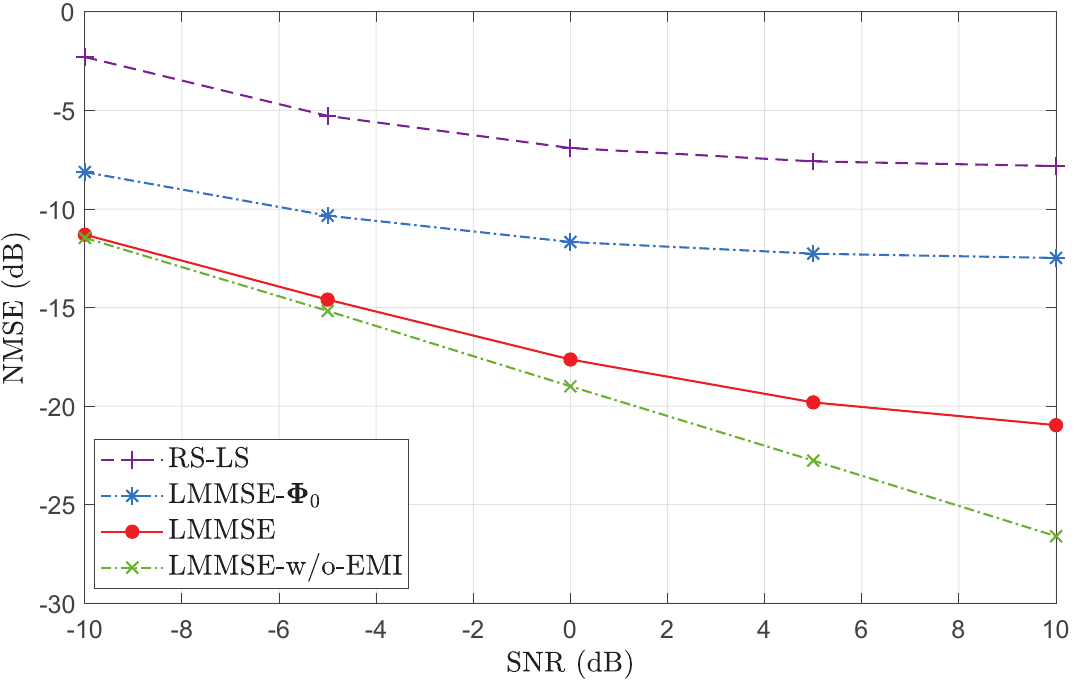}\label{Fig1a}}
\subfigure[Carrier frequency $f_0=3\THz$.]{\includegraphics[width=7.9 cm]{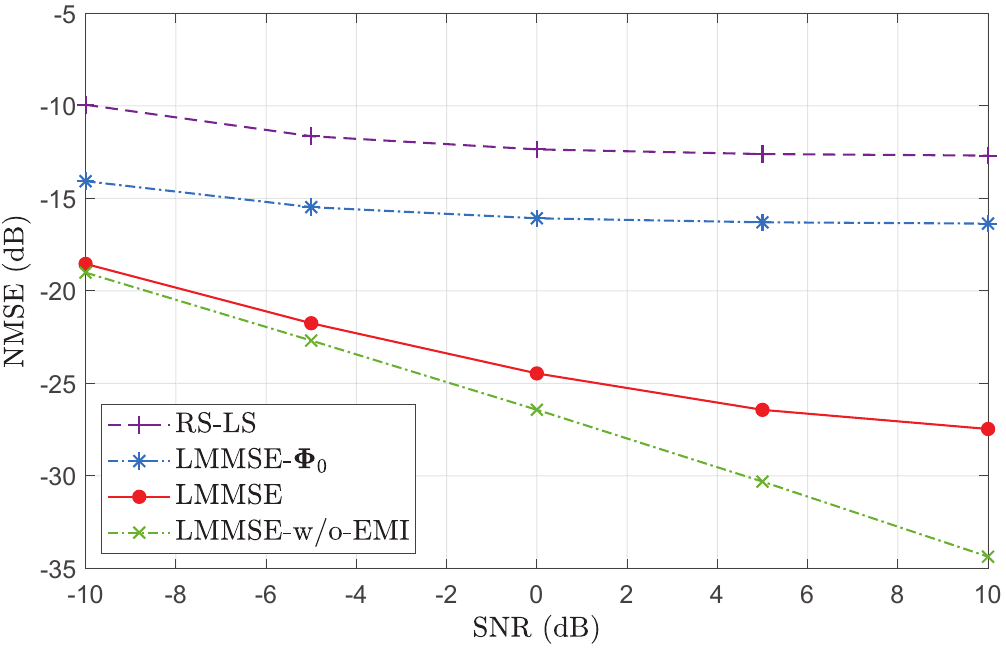}\label{Fig1c}}
\end{center}
\caption{\acs{NMSE} vs. \acs{SNR} for different estimators.}
\label{Fig1}
\end{figure}

\vspace{-0.4cm}
\subsection{Verification of AO Convergence}

\figurename~\ref{Fig7} depicts the \ac{NMSE} of the \ac{LMMSE} estimator with \ac{AO} as a function of the iteration index $k$ under various \acp{SIR}. The abbreviation `RS-LS Init.' refers to \ac{AO} initialized with the optimal phase-shift $\mathbf{\Phi}_{\tau}^{(0)}$ obtained from the \ac{RS-LS} estimator, while `Random Init.' denotes initialization using random unit-modulus phase-shifts. As expected, the \ac{NMSE} decreases monotonically with increasing $k$ and eventually converges to a steady value, thereby validating the convergence of the \ac{AO} algorithm. Furthermore, \ac{AO} initialized with $\mathbf{\Phi}_{\tau}^{(0)}$ exhibits a slightly faster convergence rate compared to the random initialization. Based on this observation, the \ac{RS-LS}-derived optimal phase-shift is used for initialization in all subsequent simulations to improve convergence efficiency.

\vspace{-0.4cm}
\subsection{Channel Estimation Accuracy}

\figurename~\ref{Fig1a} shows the \ac{NMSE} of various estimators versus \ac{SNR} when the \ac{SIR} is $5\dB$. The pilot length is set to $\tau = \text{rank}({\mathbf{R}_{g_m}\odot\mathbf{R}_{h}})=15$, the minimum required by the \ac{RS-LS} estimator \cite{Long2023Channel}. For comparison, we also include the \ac{LMMSE} estimator in an interference-free setting (denoted `-w/o-EMI'). As expected, the \ac{NMSE} decreases for all estimators when increasing the \ac{SNR}. In the presence of \ac{EMI}, when \ac{SNR} is below \ac{SIR}, \ac{NMSE} improves with \ac{SNR}, but saturates once \ac{SNR} $\geq$ \ac{SIR}, showing that \ac{SIR} limits performance. Additionally, although both `RS-LS' and `LMMSE-${\mathbf{\Phi}_0}$' use the same \ac{RIS} setup, the \ac{LMMSE} achieves about a $5$-dB gain in \figurename~\ref{Fig1a} due to its criterion. More importantly, the \ac{AO}-based \ac{LMMSE} estimator consistently outperforms `LMMSE-${\mathbf{\Phi}_0}$' by at least $3.3\dB$, validating the effectiveness of the \ac{AO} strategy in the \ac{NF} case. Subsequently, \figurename~\ref{Fig1c} presents the performance of various estimators at $f_0 = 3\THz$. To ensure that the specific devices and \ac{EMI} all lie within the \ac{NF} region of the \ac{RIS}, the number of \ac{RIS} elements is increased to $N=36 \times 4$. Under this setting, all estimators exhibit performance trends consistent with those observed in \figurename~\ref{Fig1a}, thereby confirming the broad applicability of the \ac{AO}-based \ac{LMMSE} method in \ac{THz} \ac{NF} scenarios.

\vspace{-0.4cm}
\subsection{Impact of System Parameters}

\begin{figure}[t!]
\setlength{\abovecaptionskip}{-0.2cm}
\setlength{\belowcaptionskip}{-0.4cm}
\begin{center}
\includegraphics[width = 7.9 cm]{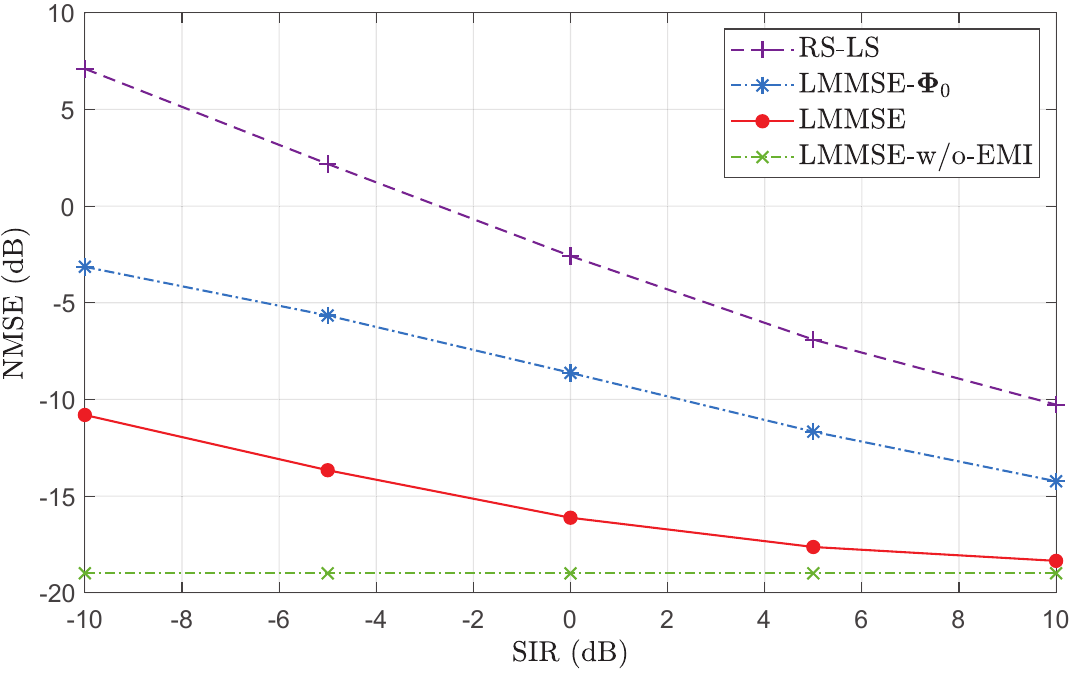}
\end{center}
\caption{\acs{NMSE} vs. \acs{SIR} for different estimators.}
\label{Fig2}
\end{figure}

\figurename~\ref{Fig2} shows the \ac{NMSE} as a function of \ac{SIR} when $\textrm{SNR}=0\dB$. As observed earlier, the \ac{AO}-based \ac{LMMSE} estimator achieves the best performance and gradually approaches the `LMMSE-w/o-EMI' baseline as \ac{SIR} increases.

\begin{figure}[t!]
\setlength{\abovecaptionskip}{-0.2cm}
\setlength{\belowcaptionskip}{-0.4cm}
\begin{center}
\includegraphics[width = 7.9 cm]{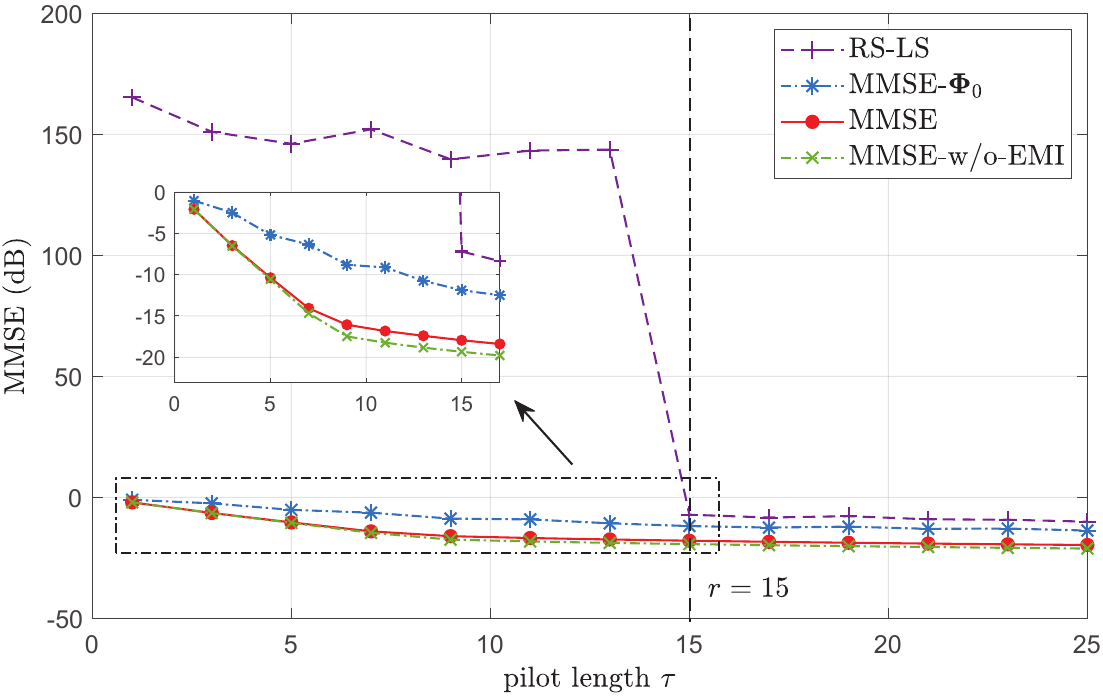}
\end{center}
\caption{\acs{NMSE} vs. $\tau$ for different estimators.}
\label{Fig3}
\end{figure}

\figurename~\ref{Fig3} presents the \ac{NMSE} as a function of the pilot length $\tau$ at $\mathrm{SNR}=0\dB$ and $\mathrm{SIR}=5\dB$, with $\tau$ ranging from $1$ to $N$. In the figure, we mark $r=\text{rank}\{\mathbf{R}_{g_m}\odot\mathbf{R}_{h}\}=15$ as the minimum pilot length required by the RS-LS estimator. The LMMSE estimator supports any $\tau\geq1$, with the most significant performance gain observed around $\tau =r$, and gradual saturation as $\tau$ approaches $N$. In contrast, the \ac{RS-LS} estimator fails when $\tau<r$ due to insufficient pilots, and its performance also saturates for $\tau\geq r$.

\begin{figure}[t!]
\setlength{\abovecaptionskip}{-0.2cm}
\setlength{\belowcaptionskip}{-0.4cm}
\begin{center}
\includegraphics[width = 7.9 cm]{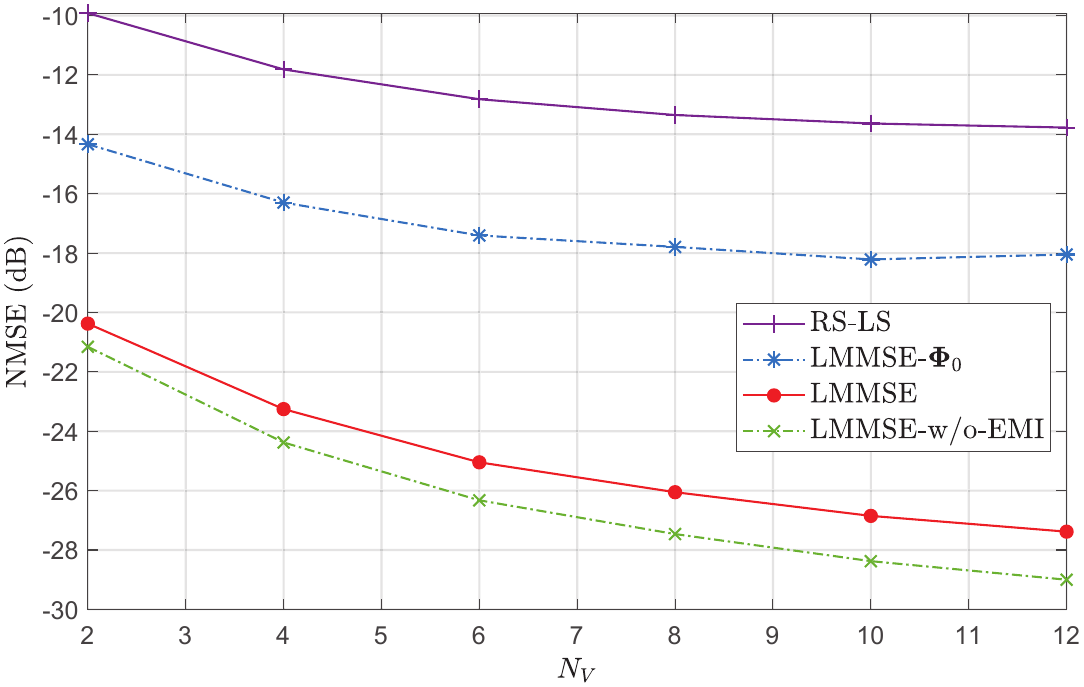}
\end{center}
\caption{\acs{NMSE} vs. $N_V$ for different estimators.}
\label{Fig4}
\end{figure}

\figurename~\ref{Fig4} depicts the \ac{NMSE} as a function of the number of \ac{RIS} elements $N$ at $\mathrm{SNR}=0\dB$ and $\mathrm{SIR}=5\dB$, where $N_H$ is fixed at $12$, while $N_V$ varies from $2$ to $12$. Note that increasing $N$ also increases the required pilot length for channel estimation. However, the impact of increasing $N$ on the cascaded channel rank, i.e., the effective channel dimension, is limited. Therefore, the dominant effect of a larger \ac{RIS} is the enhancement of the \ac{RIS} gain, which improves \ac{NF} channel estimation. Accordingly, the \ac{NMSE} of all estimators decreases as $N$ increases. It is also worth noting that the performance gap between `LMMSE' and both `RS-LS' and `LMMSE-${\mathbf{\Phi}_0}$' widens with increasing $N$. The results in \figuresname~\ref{Fig1}-\ref{Fig4} clearly demonstrate the effectiveness of the \ac{AO}-based \ac{LMMSE} estimator for \ac{NF} cascaded channel estimation in the \ac{THz} band.

\vspace{-0.4cm}
\subsection{Impact of \acs{NF} and \acs{FF} Statistical Characteristics}

\begin{figure}[t!]
\setlength{\abovecaptionskip}{-0.2cm}
\setlength{\belowcaptionskip}{-0.4cm}
\begin{center}
\includegraphics[width = 7.9 cm]{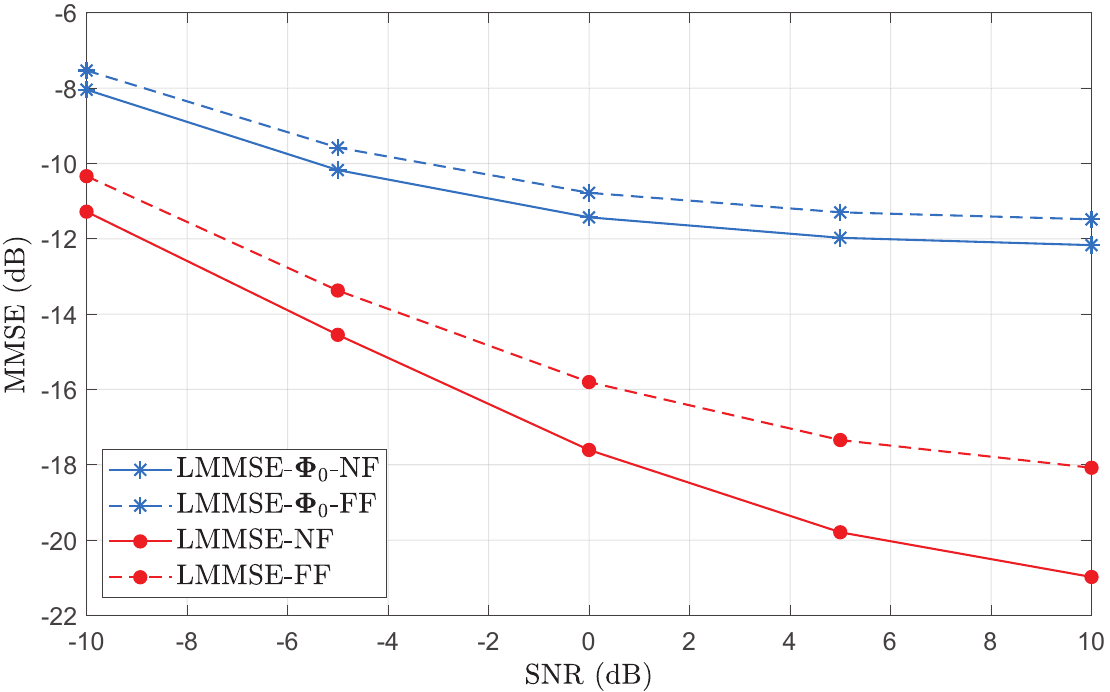}
\end{center}
\caption{\acs{NMSE} vs. \acs{SNR} for different estimators using \acs{FF} and \acs{NF} statistics in the \acs{NF} region of \acs{RIS}.}
\label{Fig5}
\end{figure}

\figurename~\ref{Fig5} shows the \ac{NMSE} of the \ac{LMMSE} estimator within the \ac{NF} region of the \ac{RIS}, based on the \ac{NF} statistics jointly characterized by distance and angles as described in \eqref{RhNF}, and the \ac{FF} statistics characterized solely by angles as given in \cite[Eq.~(3)]{Long2024MMSE}. As expected, the \ac{LMMSE} estimator based on \ac{NF} statistics consistently outperforms its \ac{FF}-based counterpart across all \ac{SNR} levels. Moreover, as the \ac{SNR} increases, the \ac{AO}-based \ac{LMMSE} estimator with \ac{NF} statistics exhibits a steeper reduction in \ac{NMSE}, yielding greater performance gains over the alternative approaches.

\begin{figure}[t!]
\setlength{\abovecaptionskip}{-0.2cm}
\setlength{\belowcaptionskip}{-0.5cm}
\begin{center}
\includegraphics[width = 7.9 cm]{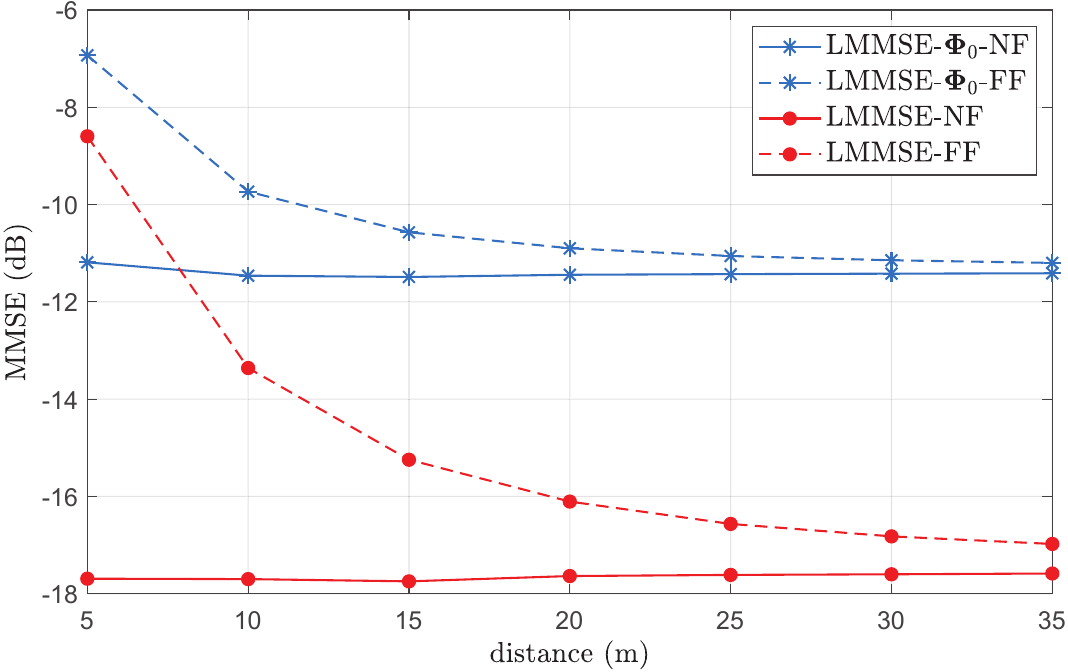}
\end{center}
\caption{The \acs{NMSE} vs. distance for different estimators using \acs{FF} and \acs{NF} statistics in the \acs{NF} region of the \acs{RIS}.}
\label{Fig6}
\end{figure}

\figurename~\ref{Fig6} presents the \ac{NMSE} of the \ac{LMMSE} estimator as a function of the distance from the specific devices and interference source to the \ac{RIS}, ranging from $5\meter$ to $35\meter$, under \ac{NF} and \ac{FF} statistical assumptions. Within the \ac{NF} region of the \ac{RIS}, i.e., distances less than $29.6\meter$, \ac{NF}-based estimators consistently outperform their \ac{FF}-based counterparts. As the distance increases, the performance of the \ac{FF}-based estimators gradually approaches that of the \ac{NF}-based ones and eventually stabilizes beyond the \ac{RIS} Rayleigh distance. Together with the results in \figurename~\ref{Fig5}, these findings underscore the importance of incorporating \ac{NF} statistics for accurate \ac{AO}-based \ac{LMMSE} estimation in the \ac{NF} region of the \ac{RIS}.

\vspace{-0.2cm}
\section{Conclusion}

We investigated the issue of cascaded channel estimation in \ac{RIS}-aided \ac{THz} wireless communications under the random \ac{EMI}. Unlike prior studies~\cite{Long2024MMSE}, we considered a scenario in which the specific devices and interference source are all located within the \ac{NF} region of the \ac{RIS}. By leveraging known \ac{NF} statistical characteristics, the \ac{AO} strategy can effectively enhance the performance of the \ac{LMMSE} estimator in \ac{NF} scenarios through the alternating optimization of the estimator and the \ac{RIS} phase-shifts. Numerical results confirmed the necessity of incorporating \ac{NF} statistics for channel estimation in such a scenario, and demonstrated the clear advantage of the \ac{AO}-based \ac{LMMSE} estimator over existing alternatives in \ac{NF} cascaded channel estimation. The acquired channel estimates subsequently provide essential information for the near-field beamfocusing at the \ac{RIS} under \ac{EMI} during data transmission.

\bibliographystyle{IEEEtran}
\bibliography{IEEEabrv,reference}

\end{document}